\documentclass[useAMS,usenatbib]{mn2e}
\usepackage{graphicx}
\usepackage{txfonts}
\usepackage{natbib}

\usepackage[english]{babel}
\usepackage{color}
\DeclareGraphicsExtensions{.pdf,.ps,.jpg}

\newcommand{\src}{SDSSJ143244.91+301435.3}

\title[SDSSJ143244.91+301435.3 at VLBI]{SDSSJ143244.91+301435.3 at VLBI: a compact radio galaxy 
in a narrow-line Seyfert 1}
\author[Caccianiga et al.]{A. Caccianiga$^1$, D. Dallacasa$^{2,3}$,
S. Ant\'on$^{4}$, L. Ballo$^1$, M. Berton$^5$, K.-H. Mack$^3$,  
\newauthor A. Paulino-Afonso$^4$ \\
\vspace{0.2cm}
\\
   $^1$INAF - Osservatorio Astronomico di Brera, via Brera 28,  I-20121 Milan, Italy\\
   $^2$Dipartimento di Astronomia, Universit\`a di Bologna, via Ranzani 1, 40127, Bologna, Italy\\
   $^3$INAF - Istituto di Radioastronomia, Via Gobetti 101, I-40129 Bologna, Italy\\
   $^4$Instituto de Astrof\'sica e Ci\^encias do Espa\c{c}o, Universidade de Lisboa, Faculdade de Ci\^encias, 
Campo Grande, PT1749-016 Lisboa, Portugal\\
   $^5$Dipartimento di Fisica e Astronomia ``G. Galilei'', Universit\'a di Padova, Vicolo dell'Osservatorio 3, 35122, Padova, Italy
 }
   \date{}

\begin{document}

\label{firstpage}

\maketitle

\begin{abstract}
We present VLBI observations, carried out with the European Very Long Baseline
Interferometry Network (EVN), of \src, a radio-loud narrow-line Seyfert 1 (RL~NLS1) characterized by a steep radio spectrum. 
The source, compact at Very Large Array (VLA) resolution,  
is resolved on the milliarcsec scale, showing a central region plus two extended structures.
The relatively
high brightness temperature of all components (5$\times$10$^6$-1.3$\times$10$^8$ K) supports the hypothesis that the
radio emission is non-thermal and likely produced by a relativistic jet and/or small radio lobes. 
The observed radio morphology, the lack of a significant core and the presence of a low frequency (230 MHz)
spectral turnover are reminiscent of the 
Compact Steep Spectrum sources (CSS). However, the linear size of the source ($\sim$0.5~kpc) measured from the
EVN map is  lower than the value predicted using the turnover/size relation valid for CSS sources ($\sim$6 kpc).  
This discrepancy can be explained by an additional component not detected 
in our observations, accounting for about a quarter of the total source flux density, combined to projection effects.
The low core-dominance of the source (CD$<$0.29) confirms that \src\ is not a blazar, i.e. the relativistic jet
is not pointing towards the observer. This supports the idea that \src\ may belong to the ``parent population'' 
of flat-spectrum RL~NLS1 and favours the hypothesis of a direct link between RL~NLS1 and compact, possibly 
young, radio galaxies.

\end{abstract}

\begin{keywords}
galaxies: active -  galaxies: nuclei - quasars: individual: SDSSJ143244.91+301435.3
\end{keywords}

   \maketitle


\section{Introduction}
Narrow-line Seyfert 1 (NLS1) galaxies represent a class of Active Galactic Nuclei (AGN)
characterized by narrow ($<$2000 km s$^{-1}$) Balmer lines,
weak [OIII]$\lambda$5007\AA\ emission compared to the 
Balmer lines ([OIII]$\lambda$5007\AA/H$\beta$ flux ratio below 3) and
strong optical FeII emission (e.g. Osterbrock \& Pogge 1985; Goodrich 1989; Pogge 2000; 
Veron-Cetty et al. 2001). The optical 
properties of NLS1 are usually explained as the consequence of a 
combination of a low-mass  ($\lesssim$10$^8$ M$_{\odot}$)  
central supermassive black-hole (SMBH)
with a high accretion rate (close to the Eddington limit).
Most of the NLS1 are radio-quiet (RQ) objects while a minority 
($\sim$7 per cent, \citealt{Komossa2006}) 
are radio-loud (RL), using the common definition based
on the 5~GHz to 4400\AA\ flux density ratio (R$_5$, where RL NLS1 have R$_5>$10)
or an (almost) equivalent
definition based on the 1.4~GHz flux density 
(R$_{1.4}$, where RL NLS1 have R$_{1.4}>$19, \citealt{Komossa2006}).
The $\sim$110 RL~NLS1 discovered so far, 
have been discussed in a number of papers (e.g. \citealt{Komossa2006}; 
\citealt{Yuan2008}; \citealt{Foschini2015}; \citealt{Berton2015}; \citealt{Gu2015}).

A peculiarity of the RL NLS1 discovered to date is 
that most of them present a compact radio emission (linear size below 
a few kpc, e.g.  Doi et al. 2012 and  references therein). 
Some of the RL NLS1 with the highest  values of radio-loudness show strict
similarities with blazars (BL Lac objects 
and flat spectrum radio quasar, FSRQ): a flat or inverted radio spectrum, high 
brightness temperatures ($T_B>$10$^{11}$ K, e.g. 
\citealt{Yuan2008}) and 
a detectable gamma-ray emission (in {\it Fermi}-LAT, \citealt{Abdo2009}; 
\citealt{Abdo2009a}; \citealt{Foschini2011}; \citealt{Foschini2015}; \citealt{Liao2015}; 
\citealt{Yao2015}; \citealt{Yao2015a}). 
Since blazars are usually believed to
be radio galaxies whose relativistic jets are pointing towards the observer 
(e.g. \citealt{Urry1995}), 
a natural conclusion is that most of the 
RL NLS1 discovered so far 
should also have their jets aligned within small angles to the line of sight.
If  this picture is correct we expect a population of mis-oriented and 
unbeamed sources, the so-called {\it parent population}, that in the standard 
beaming model (e.g. \citealt{Urry1995}) is constituted by the class of lobe-dominated radio galaxies.
To date, however, only in few RL NLS1 an extended 
emission has been detected (\citealt{Whalen2006}; \citealt{Anton2008}; 
\citealt{Doi2012}; \citealt{Richards2015}) and only one RL NLS1  (SDSSJ120014.08--004638.7) is a 
lobe-dominated radio galaxy (\citealt{Doi2012}). 
This may suggest that RL NLS1 are intrinsically lacking an extended radio emission component 
on large scales or that these structures are not detected due to selection effects
(e.g. see \citealt{Richards2015}).
Several hypotheses for the parent population have been recently considered and discussed
by \citet{Berton2015} and \citet{Berton2016c}.  

An interesting possibility is that RL NLS1 are young or ``frustrated''  
radio galaxies that either have not yet fully deployed their radio lobes 
on large scales (at least tens of kpc) or they will never be able to form them.
This possibility was suggested on the basis of the similarities found between 
some RL NLS1 and Compact Steep-Spectrum (CSS)  or  GHz-Peaked Spectrum 
(GPS) sources (e.g. \citealt{Oshlack2001}; \citealt{Gallo2006};
\citealt{Komossa2006}; \citealt{Yuan2008}; \citealt{Caccianiga2014b}; 
\citealt{Gu2015}; \citealt{Gu2016}; \citealt{Schulz2016}) 
that are usually believed to be young radio galaxies 
(e.g. \citealt{Fanti1995}).
 
A direct test of this hypothesis, however, is not simple. In sources with the jet oriented close
to the line of sight, i.e. RL NLS1 with blazar properties, the analysis of any 
possible extended emission is difficult due to the dominance of the
beamed nuclear emission.
Conversely, if the sources have the jet axis close to the plane of the sky, 
the possible presence of obscuration in the optical and X-rays makes it difficult, 
if not impossible, to study the nuclear properties and, therefore, to
establish the NLS1 nature. A direct test would only be possible by finding and
studying sources oriented at intermediate angles, where the NLS1 nature
can be established and, at the same time, the amplification of the nuclear
emission is not too severe for the analysis of the extended radio properties to be
fruitfully carried out. 
We believe that we have found at least one of such objects: \src,
a new RL NLS1 that we have recently discussed (\citealt{Caccianiga2014b}, hereafter referred 
as C14) and which shows radio properties suggesting a non-blazar nature. 
In this second paper we present the results of a radio follow up 
at a resolution of $\sim$0.01 arcsecond carried out 
using the European VLBI Network (EVN) at 1.66~GHz.

Throughout the paper  spectral indices are given assuming S$_{\nu}\propto\nu^{-\alpha}$, and  we assume a flat $\Lambda$CDM cosmology with H$_0$=71 km 
s$^{-1}$ Mpc$^{-1}$, $\Omega_{\Lambda}$=0.7 and $\Omega_{M}$=0.3.

\section{SDSSJ143244.91+301435.3: a rare case of steep-spectrum RL NLS1}

   \begin{figure}
   \centering
    \includegraphics[width=8cm, angle=0]{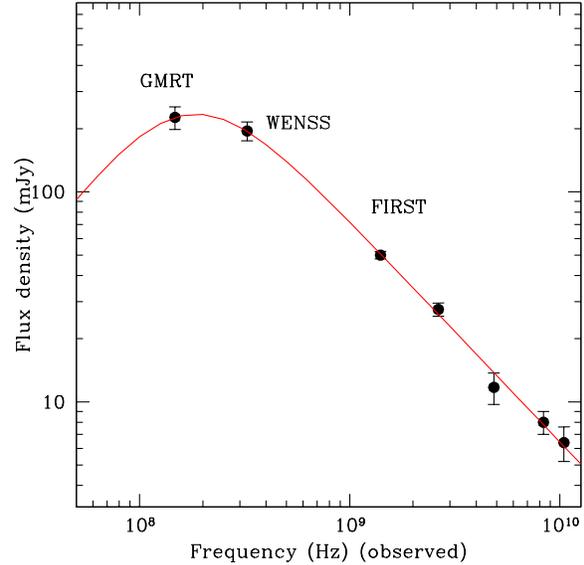}
   \caption{Radio spectrum of \src. 
High-frequency data ($>$2 GHz) come from dedicated 
observations at the Effelsberg radio telescope 
while lower-frequency data come from existing 
catalogues (see labels, more details in C14). 
Points are fitted with a smooth broken power law (red line) 
which shows a turnover at 170~MHz.
}
              \label{radio_spect}
    \end{figure}

SDSSJ143244.91+301435.3  (z=0.355, implying a conversion factor of 
4.9 pc/m.a.s.) 
is one of the few (a dozen in total discovered so far)  
RL NLS1 with a radio-loudness parameter, R$_{1.4}$, of the order of 500 
(corresponding to $R_{5}$=120, computed at 5~GHz).
The radio emission, however, 
is different from that usually observed in RL NLS1 with such a high value of 
R$_{1.4}$, 
since it shows a steep ($\alpha$=0.93) optically thin radio spectrum
(Fig.~\ref{radio_spect}).
Based on the size of the radio emission ($\leq$1.4 kpc, from the analysis of 
FIRST data) this object is a compact radio source. 
The steep radio spectrum and the lack of variability and polarization 
(as discussed in C14), 
instead, disfavor the hypothesis that the observed compactness is due to 
the orientation of a relativistic jet towards the observer, as in blazars. 
The compact morphology and the steep radio spectrum are characteristics 
remarkably similar to those observed in CSS sources. 
The radio spectrum shows a flattening below 300 MHz,
consistent with the presence of a turnover around 200~MHz, a distinctive 
feature observed in CSS sources. However the lack of sensitive data at 
low frequencies, at the time of the publication of C14, prevented us from 
firmly establishing the position of such possible turnover. 

Since the publication of C14 a new sensitive all-sky survey
at 150~MHz carried out with
the Giant Metrewave Radio Telescope (GMRT) has been 
released (\citealt{Intema2016}). Thanks to the good sensitivity of this survey,
\src\ is clearly detected and its flux density is well determined.  
The total and the peak flux densities (S$_{150 MHz}$=226.1$\pm$27.9 mJy, 
S$^{peak}_{150 MHz}$=206.8$\pm$23.8 mJy/beam) are not significantly
different thus confirming the compactness of the
source also at this low frequency (the resolution is 
25$\arcsec\times$25$\arcsec$). In C14 we used 
the flux density
taken from the 7C survey (\citealt{Hales2007}) which, however, is more
uncertain (330 $\pm$80 mJy) than the GMRT measurement and is likely to 
include some contribution from nearby sources (see C14).
We now use  the GMRT flux density and study again the shape of 
the radio spectrum 
(Fig.~\ref{radio_spect}). 

We fit the observed points with a smooth broken
 power-law model:

\begin{equation}
 S (\nu) = \frac{ 2 S_{0}} { (\frac {\nu}{\nu_{0}})^{\alpha} + (\frac {\nu}{\nu_{0}})^{\beta} }
\label{bpl.eq}
\end{equation}

\begin{table}
\begin{center}
\begin{tabular}{c c c c c}
\hline
Log $S_0$      & Log $\nu_0$    & $\alpha$            & $\beta$              & Red. Chi. sq \\
(mJy)        & (Hz)         &               &                &              \\ 
\hline
2.37         & 8.23         & 1.05          & -1.28          &    1.26      \\
(2.33, 2.41) & (8.17, 8.33) & (1.00, 1.11)  & (-3.03, -0.45) &             \\
\hline
\end{tabular}
\end{center}
\caption{Best-fit parameters of the spectral fit with a smooth broken power-law (see text for details). 
Error intervals (68\% confidence level) on the parameters are given below the best-fit values.}
\label{bestfit}
\end{table}

The best fit parameters are reported in table~\ref{bestfit}. 
The turnover at 170~MHz (observed frame) corresponds to an intrinsic turnover at
230 MHz (rest frame). 
The value of the turnover is within the range observed in CSS 
(e.g. \citealt{Fanti1995}; \citealt{O'Dea1998}). 
A more accurate characterization of the turnover will be
possible in the near future 
thanks to the availability of the data from the ongoing Multifrequency Snapshot Sky 
Survey (MSSS) carried out with LOFAR in the 30 to 74 MHz (LBA) and in the 120 to 160 MHz (HBA)
frequency range. 

A second important characteristic of \src\ is the presence of a highly star-forming 
host galaxy (star-formation rate, SFR, $\sim$50 M$_{\sun}$ y$^{-1}$), as inferred from 
the analysis of the optical-to-mid-IR Spectral Energy Distribution (see C14). 
This characteristic seems
to be common to most of the RL~NLS1 discovered so far, as discussed in \citet{Caccianiga2015}.
Since star-formation (SF) activity is expected to produce a significant radio emission, it
is possible that the radio flux density observed in some RL NLS1 is not entirely due to the AGN  
(see discussion in \citealt{Caccianiga2015}).

Overall, the properties of \src, and in general of RL~NLS1, suggest a complex
phenomenology that require high resolution radio data to distinguish between
the different components (e.g. jet vs SF) and to establish the possible connection with other classes 
of radio sources, like the CSS/GPS radio galaxies.  
With these goals we have carried out a follow-up of \src\ 
using the EVN. 
The results are discussed in the following sections.

\section{EVN observation}

   \begin{figure*}
   \centering
      \includegraphics[width=12cm, angle=0]{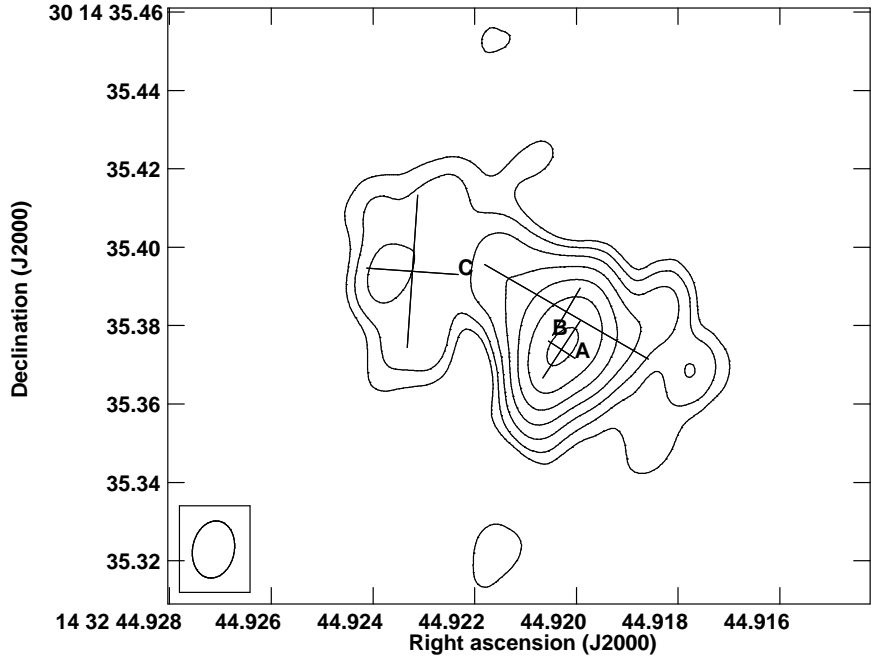}
 \caption{EVN map at 1.658~GHz of SDSSJ143244.91+301435.3. The beam size is 
14.7$\times$10.6 m.a.s. in p.a. $-11^\circ$ and it is shown in the inset at the bottom-left corner, while the rms of the map is $\sim$50 $\mu$Jy/beam. 
Levels correspond to -1,1,2,4,8,16,32,64 times 0.148 mJy/beam (=3$\times$rms). 
At the redshift of the source 1 m.a.s. corresponds to 4.9 pc.
}
              \label{evn_map}
    \end{figure*}

\subsection{Observations and data reduction}

\src\ was observed at 1.658~GHz with the EVN from UT=20:00 June 01, to UT=02:00 June 02, 2014 (project EC048).
Ten telescopes participated in the observation, but only eight provided useful data, namely Effelsberg (Germany), 
Noto (Italy), Onsala (Sweden), Toru\'n (Poland), 
Shanghai (China), the Westerbork Synthesis Radio Telescope 
(WSRT, the Netherlands), Svetloe and  Zelenchukskaya (Russia) 
while Badary (Russia) and Jodrell Bank (UK) could not provide any useful data due to
technical problems. The JIVE correlator delivered data with 2s integrations
for parallel hand data (RR and LL correlations) for 8 IFs covering
8MHz bandwidth each, for a total bandwidth of 64 MHz for each polarisation.

The phase-referenced observation of the target was interleaved with the observation of 
a phase calibrator (J1435+3012) which is 0.61 degrees apart from \src, with a duty cycle in which  the telescopes tracked the calibrator and target for 2 and 7.5 min respectively.
Four 2-min scans on an additional calibrator (J1635+3808) were carried out for calibrating the amplitudes and the bandpass profile.

The EVN data were then processed using  the NRAO {\it AIPS} package, following standard procedures of amplitude  and bandpass calibration, fringe fitting of the calibration sources and solution transfer to the target, as in phase-referencing mode observation data reduction, allowing the recovery of the absolute position of the radio emitting components on the sky. Based on gain variations during the experiment we conservatively assume that the accuracy of the calibration of the absolute flux-density scale is within 10 per cent. 
Then  accurate imaging was performed for all the sources to test the quality of the a-priori calibration, including 
self-calibration. No significant problem was found. 

The field of the target source presents significant emission with the peak shifted by about 50 m.a.s. in RA and -140 m.a.s. in Dec, with respect to the position used for the correlation (RA= 14:32:44.917; Dec= 30:14:35.510). The final uniformly weighted, full-resolution image has a restoring beam of 14.7 x 10.6 m.a.s. in p.a. $-11^\circ$, and a r.m.s. noise level of 
50~$\mu$Jy/beam. Since the image accounts for about 34 mJy, a few iterations of phase-only self-calibration were performed.
We also obtained a naturally weighted image (not shown) with a restoring beam of 27.1 x 18.3 m.a.s. in p.a. $-18^\circ$ accounting for exactly the same total flux density.

\subsection{Analysis}
The full-resolution EVN map (Fig.~\ref{evn_map}) shows 
a rather complex structure, without a straightforward classification of the morphology.
The peak in the image is 10.75 mJy/beam at RA=14h32m44.9203s and DEC=30d14\arcmin35.374\arcsec. 
We performed a number of Gaussian fits (task JMFIT) to the central region accounting for about 34 mJy  varying the size of the fitted region and the starting model.  The  best results were obtained with  three components (see Tab~1) labeled A, B and C in Fig.~\ref{evn_map}, where the asymmetric crosses highlight the major, minor axes and the position angle of the major axis of each component.
The best fit was chosen on the basis of the residual image and its r.m.s obtained after the subtraction of the model image from the one in Fig.~\ref{evn_map}. Indeed the fitting of all the three components  simultaneously did not provide a fair result. 
Our best result was obtained by selecting two regions to fit components A+B, first and then another region for fitting C alone. The total flux density of the three components exceeds by about 1 mJy the measurement taken by means of the task TVSTAT. A fit with two Gaussian components, roughly coincident with A and B, was in better agreement with the TVSTAT flux density, but left some residual flux density in the East part of the radio emitting region and provided a Gaussian profile more elongated than that reported in table 2. 

We extrapolated the source flux density at 1.66
GHz by considering the flux density of 50 mJy measured by both the FIRST (\citealt{Becker1995}) and
the NVSS (\citealt{Condon1998}) surveys at 1.46 GHz and at lower resolution, scaling it by the
optically thin spectral index (see Section 2). This yields a predicted total
flux density of about 44 mJy.
Therefore, about 10 mJy (22 per cent of the total flux density) are not accounted for in our observation.

{\bf Component A} accounts for the brightest region in Fig.~\ref{evn_map}. It turns out to be resolved, with an axial ratio of about 2.2, with the de-convolved major axis larger than the beam size and the minor axis exceeding the 70\% of the beam in its direction. It is likely that this region harbours the core, which would be better visible at higher frequencies and with better resolutions.
As discussed in section 2, the overall spectrum of the source is quite steep, and therefore the absence of a dominant core is not surprising. The whole region A accounts for about $\sim$ 45 per cent of the total flux density of the source, and the true core may account at most for a few percent of the total flux density at 1.66~GHz, in agreement with what is known in CSS-GPS radio sources, where the cores are generally weak (e.g. \citealt{Dallacasa2013}).
{\bf Component B} roughly describes the largest emitting region in Fig.~\ref{evn_map}, elongated from the NE to the SW, with an axial ratio of 3.3. 
Finally, component C has the lowest surface brightness detected here and a roundish structure, with an axial ratio of 1.7.

It is not clear whether the outer regions (extending on both sides of component A) can be considered as asymmetric mini lobes of a young radio loud AGN. This is often found in CSS and GPS sources (e.g. \citealt{Orienti2007}) The NE region appears rather broad and the jet interpretation can be ruled out. The SW region is closer to the central component and we would need a better resolution to properly define its nature. It is however consistent with being the opposite counterpart of the NE emission.
For a firm characterisation of the properties of the source at VLBI scales, images at another frequency are necessary. At the moment we can exclude a core-jet structure and support a morphology consistent with young radio sources.

Since we have no evidence for variability in this source (see discussion
in C14)
the 
$\sim$10 mJy missing flux density in our image are likely to be associated with an extended component with 
a low surface brightness, below the detection threshold of our EVN image. 

A useful parameter to quantify the relative importance of the core with respect 
to the extended emission is the core dominance (CD) which 
is defined as the ratio between the core and the 
extended emission in a radio source. If we use the peak flux density measured
from the EVN map (S$_{pk}$=10.75 mJy/beam) as an upper limit of the core 
emission and the extrapolated
FIRST/NVSS flux density as the total flux density (S$_{tot}$=44 mJy), the value of 
CD is:

\begin{equation}
CD=S_{pk}/(S_{tot}-S_{pk})<0.32
\end{equation}

In case we consider the peak of component A (S$_{pk}$=9.96 mJy/beam), the core dominance lowers to 0.29.
The CD value indicates that \src\ is not core-dominated (CD$>$1), in agreement
with the observed steep radio radio spectrum. 

\begin{table*}
\begin{center}
\begin{tabular}{l c c c c c c c c}
\hline
Comp. & RA & DEC. &Peak flux density &Total flux density & Maj Axis & Min Axis & P.A. & T$_B$ \\
&  (J2000.0)&J(2000.0)&(mJy beam$^{-1}$)&  (mJy)             & (m.a.s.) & (m.a.s.) & (Deg) & (K) \\ 
\hline
A & 14 32 44.9203 & +30 14 35.374 &9.96 $\pm$ 0.05  & 19.70 $\pm$ 0.14  & 17.4 $\pm$ 0.2 &  8.0 $\pm$ 0.2 &  146 $\pm$ 1 & 1.26$\pm0.04 \times10^8$   \\
B & 14 32 44.9202 & +30 14 35.383 &1.76 $\pm$ 0.05  & 11.21 $\pm$ 0.34  & 48.2 $\pm$ 1.3 & 14.8 $\pm$ 0.8 &  ~64 $\pm$ 1 & 1.39$\pm0.09 \times10^7$  \\
C & 14 32 44.9232 & +30 14 35.394 &0.70 $\pm$ 0.05  & ~4.79 $\pm$ 0.36  & 38.9 $\pm$ 3.0 & 23.3 $\pm$ 1.9 &  175 $\pm$ 1 & 4.69$\pm0.63 \times10^6$  \\
\hline
\end{tabular}
\end{center}
 
\caption{Results from the analysis of the EVN map. The errors reported here for the peak and total flux density do not take
into account the uncertainty on the absolute flux density scale calibration.  The angular sizes reported here are the de-convolved values of the Gaussian component of the fit, and their uncertainty is that provided by the fitting procedure.}
 
\end{table*}

   \begin{figure*}
   \centering
    \includegraphics[width=10cm, angle=-90]{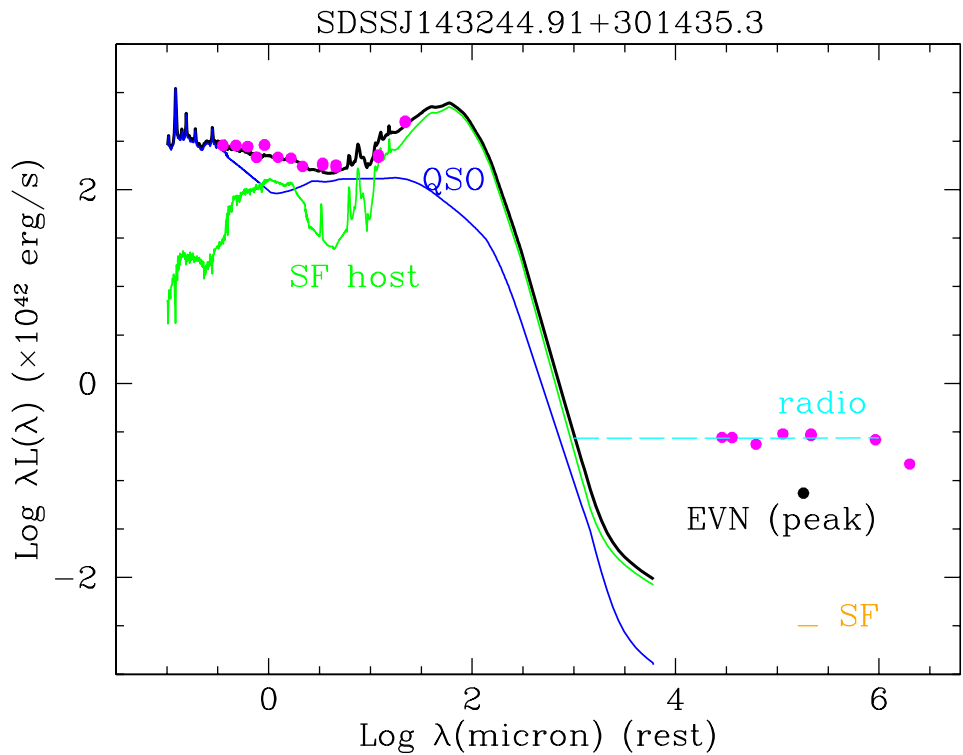}
   \caption{Spectral Energy Distribution of \src, from UV to radio. 
The photometric points (magenta in the electronic version) have been
modelled with a combination of different templates (as described in C14)
including a starburst host galaxy (M82, green line in the electronic version) plus
a mildly reddened (A$_V$=0.1 mag) QSO template (blue line in the electronic version).
It is also shown the peak emission at 1.4~GHz as measured by the EVN observations
discussed in this paper. 
Finally we show the predicted level of radio emission (at 1.4~GHz) due to the
SF, as estimated in C14 (orange short line). }
              \label{sed1432p3014}
    \end{figure*}

\subsection{Jet vs. star-formation}
In C14 we have found that \src\ is harboured by a highly
star-forming host galaxy, with an estimated star-formation rate of 50 M$_{\sun}$ y$^{-1}$. 
In principle, a fraction of the observed radio emission could be due to SF activity. 
In Fig.~\ref{sed1432p3014} we present
the Spectral Energy Distribution (SED) of \src, from UV to radio wavelengths where 
we also show the different templates that have been used to model the SED, including a radio-quiet
QSO and a starburst host galaxy template (M82, see C14 for details). 
Since the M82 template does not
include the radio band, we have added the expected emission from the SF component at 1.4~GHz
based on the analysis described in C14. Basically, this is the emission expected
from the mid-IR luminosity of the starburst host. As discussed in  C14,
the expected contribution from the SF to the radio emission is negligible ($\sim$1 per cent). 
VLBI observations can be used to directly test this conclusion, since high brightness 
temperatures (T$_B\geq$10$^7$ K) are not expected in case of thermal emission (e.g. 
\citealt{Giroletti2009}).

We compute the value of T$_B$ using the relation (\citealt{Ulvestad2005}):

\begin{equation}
T_B=1.8 \times 10^{9} \frac{(1+z) S_{\nu}}{\nu^2 \theta_{maj} \theta_{min}} K.
\end{equation}

where $S_{\nu}$ is the flux density at 1.6 GHz expressed in mJy, $\nu$ is the observed frequency in GHz, 
$\theta_{maj}\theta_{min}$ is the de-convolved size of the source in m.a.s.$^2$  and $z$ is the redshift. 
For z=0.355, $\nu$=1.658~GHz and using the flux densities and de-convolved sizes reported in Table~1 
we obtain $T_B$=1.3$\times$10$^8$ K, 1.4$\times$10$^7$ K and 4.7$\times$10$^6$ K for the 3 Gaussian components.
These values disfavour the hypothesis of a thermal origin of the
emission (e.g. from SF) and support the idea that we are observing the
non-thermal emission from jets/lobes. Further support could come from polarimetric studies
to be carried out at high frequencies, to minimize the severe depolarization 
usually observed in young radio sources (e.g. \citealt{Rossetti2008})

\subsection{Linear size}
The NE-SW elongation of \src\ on the EVN map is about 100 m.a.s. 
which translates into a (projected) linear size
of about 0.5~kpc.
Given the observed correlation between
the turnover frequency and the source linear size in CSS and GPS sources 
(\citealt{O'Dea1998, Orienti2013a}) it is possible to infer an independent 
estimate of the linear size of \src\ starting from the position of the turnover.
If we assume the turnover frequency of 230~MHz computed in Section~2 and if we  
use the relation in \citet{Orienti2013a}, the size of the source is predicted to be 
about 6~kpc, which is one order of magnitude larger than the value measured from the EVN map.
The difference is significant even considering the scatter of the turnover/size relation and 
the uncertainty on the turnover frequency.

A possible explanation for the observed discrepancy is that the EVN observations are missing the most extended 
structures of the source. 
Indeed, as previously discussed, we know that the EVN observation is missing 
some flux density ($\sim$10 mJy) which is likely distributed in extended structures,
located between 0.5 kpc and $\sim$1.4 kpc, i.e. the maximum size measured by 
the low resolution map (FIRST). However, the upper limit of 1.4~kpc based on 
the FIRST map is still below the expectations from the turnover frequency, although
the difference is less significant considering the scatter on the turnover/size relation and
the uncertainty on the turnover frequency.
The possibility that an even more extended structure has been missed also by the
FIRST map is unlikely since the measured flux density is fully 
consistent with that measured at even lower resolution by the NVSS survey
(see C14 for details). 

Another possibility is that the small size of \src\ is
due to projection effects.  
Since in \src\ we are able to directly observe the optical nucleus without 
significant obscuration (see C14), it is likely
that the source is observed at an intermediate angle between the beaming 
cone and 
the plane of the sky. At these angles, projection effects could be important 
as in CSS with a QSO-like spectrum (e.g. \citealt{Orienti2016}).
We can roughly estimate a lower limit on the jet orientation (hence an 
upper limit on the
projection effects) using the upper limit on the 
core-dominance parameter, measured in Section~3.4. In the context of the
beaming model (\citealt{Urry1995}), this quantity is
directly related to the source's orientation by the relation:

\begin{equation}
CD = f \delta^p
\end{equation}

where $f$ is the ratio between the luminosity emitted by the jet (in the 
source's frame) 
and the unbeamed luminosity, $\delta$ is the Doppler factor defined as 
$[\Gamma(1-\beta cos\theta)]^{-1}$. Here, $\Gamma$ and $\beta$ are the 
Lorentz factor and the ratio between
the bulk velocity and the light speed respectively while
$\theta$ is the angle between the observer and the direction of the
relativistic jet. 
The exponent p is equal to $\alpha$+2 for a continuous jet 
where $\alpha$ is the spectral index (see \citealt{Urry1995} for 
a detailed description of the beaming model). 

In blazars, the values of $\Gamma$ typically range from $\sim$2 to 30 
(\citealt{Urry1995}) while
$f$ is expected to assume values between 0.01 and 1 (e.g. \citealt{Urry1995}).
For RL NLS1 these parameters have not been studied in details yet but first
results on few objects seem to indicate that the Lorentz factors of 
flat spectrum RL NLS1 are similar to those observed in blazars, although
in the lower end of the distribution (e.g. \citealt{Doi2011}; \citealt{D'Ammando2012}; 
\citealt{D'Ammando2013a}; \citealt{Lister2016}; \citealt{Fuhrmann2016}). 
The analysis of the radio luminosity function (RLF) of flat-spectrum RL NLS1
and its comparison with that of CSS sources, seems to confirm that the
average Lorentz factor ($\sim$10) in RL NLS1  is within the typical distribution observed 
in blazars, while it suggests higher values for the parameter $f$, close to 1 
(\citealt{Berton2016c}). For $f$=1 and $\Gamma$=10, the expected viewing angle is
$\sim$30\degr\ corresponding to a de-projection factor of $\sim$ 2 and a de-projected linear
size of $\sim$1 kpc for the EVN structure alone.  
A reasonable estimate of the maximum value of the projection factor can
be obtained combining a high value of Lorentz factor with a low
value of $f$. Using $\Gamma$=30 and $f$=0.01 the expected de-projection factor is about 7, 
corresponding to a maximum de-projected linear size
of $\sim$3.5 kpc. This value is within the scatter of the turnover/size relation in
\citet{Orienti2013a}. Therefore, the projection effect alone could in principle explain the discrepancy between
the measured linear size and the predictions from the turnover frequency but only if  
rather extreme values for the beaming parameters are assumed.

A more realistic possibility is that the difference between the size expected from the turnover frequency and the
measured one is a combination of both the non detection of extended structures by the EVN map
and some projection effects.

Finally, we should also mention the possibility that the
size vs turnover relation, which has been derived using objects more powerful than \src,
could not be valid at lower radio powers.

\section{Comparison with other RL~NLS1}
Out of the $\sim$110 RL NLS1 discovered so far, only $\sim$25 have been studied at 
VLBI resolution, including both flat and steep spectrum objects (\citealt{Doi2011}; \citealt{Gu2015}; 
\citealt{Orienti2015a}; \citealt{Gu2016}; \citealt{Schulz2016}). 

In general, flat spectrum RL~NLS1 show single-core or core-jet morphologies with high brightness
temperatures (up to 2$\times$10$^{11}$ K) strongly suggesting the presence of a relativistically boosted
jet (\citealt{Doi2011}; \citealt{Gu2015}; \citealt{Orienti2015a}; \citealt{Schulz2016}). 
Even if high, the values of T$_B$  are on average lower than those observed in blazars, something that
may indicate lower Doppler factors when compared to blazars (i.e. larger inclinations and/or 
lower Lorentz factors, \citealt{Gu2015}). 

RL~NLS1 with steep spectra, like \src, also show unresolved cores but with
lower brightness temperatures than flat spectrum RL~NLS1 (\citealt{Gu2015}). 
Steep spectrum
sources have often (but not always) extended structures and diffuse emission on pc scales 
(up to 100 pc, \citealt{Doi2011}; \citealt{Gu2015}). These properties 
indicate that Doppler boosting is probably not significant in most of the steep spectrum 
RL~NLS1 (\citealt{Doi2011}). The steep spectrum combined with a compact size resembles the
properties of CSS, as pointed out by \citet{Doi2011} and \citet{Gu2015}, although the observed 
morphologies appear different from those usually observed in CSS that are
typically characterized by two-sided structures and where the core is rarely observed 
(e.g. \citealt{Dallacasa1995}; \citealt{Dallacasa2013}; \citealt{Orienti2016}).

The VLBI properties of \src\ are more similar to those observed in (some) steep 
spectrum RL~NLS1 than those usually seen in flat spectrum RL~NLS1. However, \src\ 
differs from the other steep spectrum RL~NLS1 observed so far, since it lacks a strong unresolved 
core and it has most of the radio emission distributed in extended structures (within 500 pc).  
From this point of view, \src\ is morphologically more similar to CSS sources than the RL~NLS1 observed so far.

The morphology observed in \src\ presents 
also some similarities with what is observed in the few radio-quiet NLS1 studied with VLBA by \citet{Doi2013}.  
In the 5 RQ~NLS1 detected at 1.7~GHz, a large fraction of the radio power comes from diffuse emission 
components distributed within the central $\sim$300 pc nuclear region, although
a high brightness temperature ($>$6$\times$10$^7$ K) core is also detected, suggesting
the presence of a jet (\citealt{Doi2013}). In particular, MRK 1239 shows a morphology  
that resembles that observed in \src\ 
but on a smaller scale ($\sim$40 pc). Another similarity is that RQ~NLS1 have often a steep 
radio spectrum (\citealt{Moran2000a}) like the one observed in \src. In spite of these similarities,
the radio-loudness parameter of the RQ~NLS1 is two orders of magnitude lower than
in \src\ (R$_5\lesssim$1, in RQ~NLS1, to be compared to 
R$_5$=160 of \src). Another
important difference is that the masses of the central SMBH of the 5 RQ~NLS1 detected at VLBA
are significantly lower (8$\times$10$^5$-1.4$\times$10$^7$ M$_{\odot}$, \citealt{Doi2013}) 
than the mass measured in \src\ (3.2$\times$10$^7$ M$_{\odot}$, C14).

\section{Summary and Conclusions}

We have presented VLBI observations carried out
with EVN of \src, a steep spectrum
RL~NLS1 that has been recently classified as {\bf a } possible CSS source 
(C14). The results can be summarized as follows:

\begin{itemize}

\item EVN observations have resolved \src\ which shows a quite 
complex structure without a clear morphology. The total 
flux density accounted for in the EVN image (34 mJy) is divided into 3 components: a central, more compact 
(but resolved), emission, containing more than half of the total correlated flux density, 
plus two extended structures. The lack of a strong unresolved 
component (core) is consistent
with what is usually observed in steep-spectrum young radio sources.

\item The high brightness temperature of the radio components detected 
in the EVN map (T$_B\sim$5$\times$10$^6$-1.3$\times$10$^8$ K)
dis-favours the hypothesis that 
star-forming activity (detected in the mid-IR band) is the main origin of
the observed radio emission and strongly supports a non-thermal origin.

\item The size of the extended structure (100-150 m.a.s., corresponding to
$\sim$0.5~kpc) is smaller than the value inferred from the linear size/turnover
relation valid for the CSS/GPS galaxies. This is likely due to
the non-detection of a low surface brightness component in the EVN map 
combined to projection effects.

\end{itemize}

Overall, the EVN observations support the idea  that \src\ is a 
CSS source with the relativistic jet
observed at larger angles than the flat-spectrum RL~NLS1
studied so far. 
The fact of observing directly the nuclear emission in
the optical, without significant obscuration, suggests that \src\ 
is probably not oriented on the plane of the sky but at intermediate angles.
VLBI observations at high frequencies would prove this hypothesis since they
are expected to detect a stronger, mildly boosted, radio core.

These results confirm that \src\ likely belongs to the parent population of the 
RL~NLS1 with a blazar spectrum and favour the idea that these sources can be directly
related to young radio galaxies. A systematic follow-up at VLBI resolution,
similar to the one discussed here, of
all the RL~NLS1 will be fundamental to test this hypothesis on a firm statistical basis.

\section*{Acknowledgments}
We thank the referee for his/her useful comments that improved the quality of the paper.
The European VLBI Network is a joint facility of independent European, African, 
Asian, and North American radio astronomy institutes. Scientific results from 
data presented in this publication are derived from the following EVN project 
code: EC048. LB acknowledges support from the Italian Space Agency (contract ASI INAF
NuSTAR I/037/12/0).

\bibliographystyle{mn2e}
\bibliography{/home/guincho/caccia/cartella_comune/NLS1_CSS}
\end{document}